\documentclass[prb,aps,12pt]{revtex4}
\usepackage{amsfonts}
\usepackage{amsmath}
\usepackage{graphicx,bm}
\usepackage{amssymb}
\usepackage{graphicx}

\DeclareMathOperator{\Tr}{Tr}

\begin{document}

\title{Wigner distribution function formalism for superconductors and collisionless
dynamics of the superconducting order parameter}
\author{M.~H.~S.~Amin$^1$, E.~V.~Bezuglyi$^2$,  A.~S.~Kijko$^2$, and A.~N.~Omelyanchouk$^2$}
\date{\today}

\affiliation{$^1~$D-Wave Systems Inc., 320-1985 W. Broadway,
Vancouver, B.C., V6J 4Y3 Canada}

\affiliation{$^2~$B.I. Verkin Institute for Low Temperature
Physics and Engineering, Ukrainian National Academy of Sciences,
Lenin Ave. 47, Kharkov 61103, Ukraine}

\begin{abstract}
A technique to study collisionless dynamics of a homogeneous
superconducting system is developed, which is based on Riccati
parametrization of Wigner distribution function. The quantum
evolution of the superconductiung order parameter, initially
deviated from the equilibrium value, is calculated using this
technique. The effect of a time-dependent BCS paring interaction
on the dynamics of the order parameter is also studied.
\end{abstract}

\maketitle

\section{Introduction}
 In this work we study the dynamics of the superconducting order parameter within the
 Wigner distribution function approach. The problem of nonstationary phenomena in
 superconductors has been attracting attention for a long time \cite{nonsup,tinkh}.
The general method for description of nonstationary  and nonequilibrium processes is the Keldysh
 technique for nonequilibrium real time Green's functions \cite{keld}.
The equations for superconducting Keldysh Green's functions
\cite{eli,lar1} are a set of quite complicated nonlinear
integro-differential equations, which are nonlocal in time and
space domains. These equations are considerably simplified in the
quasiclassical approximation
 by integrating Green's functions over $\xi_p=p^2/2m-\mu$. \cite{lar2}
The quasiclassical Larkin-Ovchinnikov equations are still nonlocal
in time, but are local in space. In the stationary case, these
equations transform into Eilenberger's equations \cite{eilen},
which are effective tools for solving stationary inhomogeneous
problems.

When the time-dependent processes in superconductors are
considered, three
time scales are the most essential. 
The time $t_{p}\sim \omega_{p}^{-1}$ ($\omega_{p}$ is a plasma
frequency) characterizes the scale at which the self-consistent
scheme for the electromagnetic fields $\bold{A}(\bold{r},t)$,
$\phi(\bold{r},t)$, and for the BCS pairing field
$\Delta(\bold{r},t)$ is established. The time $t_{0}\sim
\Delta^{-1}$ ($\Delta$ is the superconducting gap) is an intrinsic
time for superconductors, during which quasiparticles with energy
spectrum $\sqrt{\Delta^2+\xi_p^2}$ are formed in the
superconductor. The stage of the relaxation of a nonequilibrium
disturbance in the quasiparticle distribution is determined by the
energy relaxation time $\tau_{\varepsilon}$, which is caused by
electron-phonon inelastic processes. For conventional
superconductors, at temperature $T$, not too close to the critical
temperature $T_c$, we have a hierarchy of the characteristic
times: $t_{p}\ll t_{0} \ll \tau_{\varepsilon}$. At the time
interval $t\sim \tau_{\varepsilon} \gg t_{0}$ the superconductor's
dynamics is described by the quasiclassical Boltzman kinetic
equation for the quasiparticle distribution function together with
a self-consistent equation for $\Delta(\bold{r},t)$
(Aronov-Gurevich equations \cite{aron}). In the opposite case $t
\ll \tau_{\varepsilon}$, the dynamics of the superconducting order
parameter should be described by the quantum kinetic equation.
Considering the collisionless evolution of the superconducting
order parameter ($t \ll \tau_{\varepsilon}$), the equations for
the Keldysh Green's functions are reduced to simpler equations for
the Green's functions at coinciding times. The latter can be
transformed to the quantum kinetic equation for the Wigner
distribution function (WDF). The collisionless kinetic equation
for superconducting WDF can also be obtained starting directly
from the nonstationary generalized Hartree-Fock approximation for
the BCS pairing model \cite{kul1} (see also \cite{gal1,gal2}).

Wigner introduced a distribution function in the phase space
\cite{wig} as a quantum analog of the classical Boltzman
distributions. To study quantum transport, the Wigner-function
formalism possesses many advantages. It is extensively used for
the description of normal metal and semiconducting electron
devices whose behaviour is dominated by quantum interference
effects, e.g. for self-consistent treatment of transient response
to a change in the applied voltage \cite{fre}. In recent years,
Wigner functions are widely used in the field of quantum optics to
describe the effects of quantum coherence and superposition in
optical systems \cite{optic}. Such effects are of great interest
in qubit (quantum bit for quantum computation) investigations
\cite{MSS}.

Collisionless dynamics of the superconducting order parameter has
gained renewed attention after the discovery of the BCS-like
paired state in dilute fermionic gases \cite{fermi}. The ability
to control and change the strength of the pairing interaction in
these systems opens possibilities for new experimental
investigations of the dynamics of the order parameter. Recently,
time-dependent BCS pairing was studied theoretically in
Ref.~\cite{spivak}. The WDF technique developed in this paper
provides a useful tool for studying such problems.

In Sec.~II, following to Kulik's approach \cite{kul1}, we derive a
quantum kinetic equation for superconducting WDF in $({\bf
r},t)$-space. This equation is simplified for the case of
homogeneous state (Sec.~III) and then used to study the
collisionless dynamic of the order parameter in small
superconducting systems (Sec.~IV). The problem of the time
evolution of the order parameter, initially deviated from the
equilibrium value, is considered. It appears that on times much
smaller than $\tau_\varepsilon$, the time dependence of $\Delta$
has an oscillatory nature. Earlier, such problem was studied by
other authors using a linear response approach \cite{volkov},
assuming small deviation from equilibrium. In the present work,
these dependencies were obtained under arbitrary initial
conditions (not only small). The time dependent response of the
order parameter to a time-varying pairing potential is also
studied. A numerical method for solving equation for WDF, which is
based on Maki-Schopohl transformation \cite{maki}, is developed.

\section{Wigner distribution function formalism for the superconducting state%
}

We write the Hamiltonian of the superconductor as
\begin{equation*}
H=H_{0}+H_{1},
\end{equation*}
where $H_{0}$ includes electron interactions with external fields, the
vector potential $\mathbf{A}(\mathbf{r})$ and the scalar potential $\varphi (%
\mathbf{r})$, as well as with the pairing field $\Delta (\mathbf{r})$,
\begin{align}
H_{0} &= \!\!\sum_{\sigma =\uparrow, \downarrow}\!\int\!\!\ d\mathbf{r}%
\,\psi_{\sigma}^\dagger(\mathbf{r}) \left[ {\epsilon} -\mu+e\varphi(%
\mathbf{r})\right] \psi_{\sigma}(\mathbf{r}) -\! \int \!\! d\mathbf{r}\left[%
\Delta(\mathbf{r})\psi_{\uparrow}^\dagger(\mathbf{r}) \psi_{
\downarrow}^\dagger(\mathbf{r}) + \Delta^{\ast}(\mathbf{r})\psi_{\downarrow}(%
\mathbf{r}) \psi_{\uparrow}(\mathbf{r})\right], \\
{\epsilon} &= \frac{1}{2m}\left[ \frac{\nabla}{i} -\frac{e}{c}%
\mathbf{A}(\mathbf{r})\right]^{2}
\end{align}
(we use the system of units where $\hbar=k_B=1$). Here $\psi_{\sigma}(\mathbf{r%
})=\frac{1}{\sqrt{V}} \sum\limits_{\mathbf{p}}a_{\mathbf{p}\sigma}(t)e^{i%
\mathbf{pr}}$ is the annihilation operator of an electron with the spin $%
\sigma$, and $\mu$ is the chemical potential. The Hamiltonian
$H_{1}$ describes impurity, electron-phonons, electron-electron
etc. scattering providing processes for relaxation.

The pairing field $\Delta(\mathbf{r})$ is to be determined from the
self-consistency equation
\begin{equation}
\Delta^{\ast}(\mathbf{r}) =V_{0}\left\langle \psi_{\uparrow}^\dagger(\mathbf{%
r}) \psi_{\downarrow}^\dagger(\mathbf{r})\right\rangle,  \label{Delta}
\end{equation}
where $V_0$ is the pairing potential. The electromagnetic potentials obey
Maxwell equations,
\begin{align}
&\nabla \times \mathbf{A}(\mathbf{r}) =\frac{4\pi}{c}\mathbf{j}(\mathbf{r}),
\\
&\nabla^{2}\varphi+\frac{1} {c}\frac{\partial}{\partial t} \nabla \cdot
\mathbf{A} =-4\pi\rho(\mathbf{r}),
\end{align}
where $\rho(\mathbf{r})$ and $\mathbf{j}(\mathbf{r})$ are the
charge and current densities, respectively:
\begin{align}
\rho(\mathbf{r}) &= e\sum_{\sigma}\left\langle \psi_{\sigma}^\dagger(\mathbf{%
r})\psi_{\sigma}(\mathbf{r})\right\rangle ,  \label{rho} \\
\mathbf{j}(\mathbf{r}) &= -\frac{ie}{m}\sum\limits_{\sigma}\left\langle
\psi_{\sigma}^\dagger(\mathbf{r})\nabla\psi_{\sigma}(\mathbf{r}) -
(\nabla\psi_{\sigma}^\dagger(\mathbf{r}))\psi_{\sigma}(\mathbf{r}%
)\right\rangle -\frac{2e^{2}}{mc}\mathbf{A}(\mathbf{r})\sum\limits_{\sigma
}\left\langle \psi_{\sigma}^\dagger(\mathbf{r})\psi_{\sigma}(\mathbf{r}%
)\right\rangle .  \label{j}
\end{align}

By introducing the ``particle-hole'' (Gor'kov-Nambu) representation of the
electron creation and annihilation operators in terms of 2-vectors
\begin{align}
{A}_{\mathbf{p}}=\left(
\begin{array}{c}
a_{\mathbf{p}\uparrow} \\
a_{-\mathbf{p}\downarrow}^\dagger
\end{array}
\right) , \qquad {A}_{\mathbf{p}}^\dagger=\left(
\begin{array}{cc}
a_{\mathbf{p}\uparrow}^\dagger & a_{-\mathbf{p}\downarrow}
\end{array}
\right), \\
{\Psi}(\mathbf{r})=\left(
\begin{array}{c}
\psi_\uparrow(\mathbf{r}) \\
\psi_\downarrow^\dagger(\mathbf{r})
\end{array}
\right) , \qquad {\Psi}^\dagger(\mathbf{r})=\left(
\begin{array}{cc}
\psi_\uparrow^\dagger(\mathbf{r}) & \psi_\downarrow(\mathbf{r})
\end{array}
\right),  \label{Psi}
\end{align}
we define the matrix $\widehat{f}_{\mathbf{pq}}(t)$ in the ``particle-hole''
space,
\begin{equation}
f_{\mathbf{pq}}^{\alpha \beta }(t)=\left\langle A_{\mathbf{p}-\frac{\mathbf{q%
}}{2},\beta }^{\dagger }(t)A_{\mathbf{p}+\frac{\mathbf{q}}{2},\alpha
}(t)\right\rangle ,  \notag
\end{equation}
where angle brackets denote statistical averaging and
$\alpha,\beta=1,2$ are
the indices of the vectors $A_\mathbf{p}$. The function $%
f_{pq}^{\alpha \beta }$ is the Fourier transform of the Wigner
distribution function (WDF) $f_{\alpha \beta
}(\mathbf{p},\mathbf{r},t)$ generalized to the superconducting
case,
\begin{equation}  \label{WDF}
f_{\alpha \beta }(\mathbf{p},\mathbf{r},t)= \sum_{\mathbf{q}}e^{i\mathbf{qr}%
}\left\langle A_{\mathbf{p}-\frac{\mathbf{q}}{2},\beta }^{\dagger }(t)A_{%
\mathbf{p}+\frac{\mathbf{q}}{2},\alpha }(t)\right\rangle .
\end{equation}
Correspondingly, the components of the matrix $\widehat{f}(\mathbf{p},%
\mathbf{r},t)$ are expressed through the Nambu operators $\Psi_\alpha(%
\mathbf{r},t)$ in the Heisenberg representation as
\begin{equation}
f_{\alpha\beta} =\int d\mathbf{r}^\prime e^{-i\mathbf{p r}^\prime}
\left\langle \Psi_\alpha^\dagger (\mathbf{r}+ \mathbf{r}^\prime /2,t)
\Psi_\beta (\mathbf{r}-\mathbf{r}^\prime /2,t)\right\rangle.  \label{DefWDF}
\end{equation}
It follows from Eq.~(\ref{DefWDF}) that $f_{11}$ and $f_{22}$ are real
functions, and $f_{12}=f_{21}^{\ast }$. The self-con\-sis\-tency equations,
Eqs.~(\ref{Delta}), (\ref{rho}), and (\ref{j}) can be written in terms of $%
\widehat{f}$ as
\begin{align}
\Delta &=V_0 \int \frac{d\mathbf{p}}{(2\pi )^3} \Tr \tau_- \widehat{f}(%
\mathbf{p}),  \label{Delta1} \\
\rho &=e\int \frac{d\mathbf{p}}{(2\pi )^3}\Tr \tau_3 \widehat{f}(\mathbf{p}),
\label{rho1} \\
\mathbf{j} &=\int \frac{d\mathbf{p}}{(2\pi )^3}\, \Tr \tilde{\mathbf{p}}
\widehat{f}(\mathbf{p}),  \label{j1}
\end{align}
where $\tilde{\mathbf{p}} = \mathbf{p}-e\tau_3\mathbf{A}/c$, $\tau_- =
(1/2)(\tau_1-i\tau_2)$, and $\tau_i$ are the Pauli matrices.

The evolution equation for the WDF can be derived from the
equation of motion for the electron field operators ${\psi}
=\psi_\sigma(\mathbf{r},t)$:
\begin{equation}
i\frac{\partial {\psi }}{\partial t}=\left[ {\psi }, {H}\right].
\label{eqmot}
\end{equation}
Restricting our consideration by the collisionless stage of the
evolution, we neglect the interaction part $\widehat{H}_1$ of the
Hamiltonian and obtain from Eq.~(\ref{eqmot}) the equations of
motion for the Nambu operators ${\Psi}(\mathbf{r},t)$
\begin{equation}
\left[i\frac{\partial}{\partial t} -\tau_3 \left(\widehat{\xi}
+e\varphi \right) + \widehat{\Delta}\right]{\Psi}=0, \quad
\widehat{\Delta} = \left(
\begin{array}{ccc}
0 & \Delta &  \\
\Delta^\ast & 0 &
\end{array}
\right),  \label{eqmot1}
\end{equation}
where ${\xi} = -(\nabla+ie\tau_3\mathbf{A}/c)^2/2m-\mu$. By making
use of the definition of the WDF in Eq.~(\ref{DefWDF}), we arrive,
after
some algebra, at the following dynamic equation for $\widehat f(\mathbf{p},%
\mathbf{r},t)$
\begin{equation}  \label{eqWDF}
{\frac{\partial \widehat f }{\partial t}} + i\left[ \frac{(\tilde{\mathbf{p}}%
-i\tilde\nabla/2)^2}{2m} \tau_3, \widehat f \right] + i\left[ e
\varphi \tau_3 - \widehat \Delta , \widehat f \right]_\otimes = 0,
\end{equation}
where $[\ldots ]$ denotes usual commutator, in which we consider
$\tilde\nabla$
as an integral operator with the kernel $\nabla_{\mathbf{r}}\delta(\mathbf{r}%
-\mathbf{r}^\prime)$, thus $(\tilde\nabla \widehat f)=-(\widehat f
\tilde\nabla)= \nabla \widehat f$. The quantity $[\ldots]_\otimes$
is defined
as $[A,B]_\otimes \equiv A \otimes B - B \otimes A$, where $(A\otimes B)(%
\mathbf{p},\mathbf{r},t)$ is the Fourier transformation of the spatial
convolution $(AB)(\mathbf{r}_1, \mathbf{r}_2)=\int d\mathbf{r}\,A(\mathbf{r}%
_1, \mathbf{r}) B(\mathbf{r}, \mathbf{r}_2)$:
\begin{align}
(A\otimes B)(\mathbf{p},\mathbf{r}) &= \int\! d\mathbf{r}^\prime e^{-i%
\mathbf{p r}^\prime} (AB) (\mathbf{r}+ \mathbf{r}^\prime /2, \mathbf{r}-%
\mathbf{r}^\prime /2) = e^{{\frac{i}{2}}\left\{ \partial_{\mathbf{r}%
}^{A}\cdot \partial_{\mathbf{p}}^{B}-\partial _{\mathbf{p}}^{A}\cdot
\partial _{\mathbf{r}}^{B}\right\} }A(\mathbf{p},\mathbf{r}) B(\mathbf{p},%
\mathbf{r}).
\end{align}

By making use of the transformation $\widehat{f} \to
\exp(i\tau_3\chi/2)\widehat{f} \exp(-i\tau_3\chi/2)$, we can
exclude the phase $\chi$ of the superconducting order parameter
and proceed to gauge invariant quantities, i.e.
the momentum of the superfluid condensate $\mathbf{p}_s$ and the potential $%
\Phi$ defined by
\begin{equation}  \label{psphi}
\mathbf{p}_s = m\mathbf{v}_s = \frac{1}{2} \left(\nabla\chi - \frac{2e}{c}
\mathbf{A}\right), \quad \Phi =\frac{1}{2} \left(\frac{\partial\chi }{%
\partial t} + 2e \varphi \right).
\end{equation}
The electromagnetic fields are related to $\mathbf{p}_s$ and
$\Phi$ through
\begin{equation}  \label{EH}
e\mathbf{E} = {\frac{\partial \mathbf{p}_s }{\partial t}} - \nabla \Phi ,
\qquad e\mathbf{H} = -\nabla \times \mathbf{p}_s .
\end{equation}
This results in the substitutions $\tilde{\mathbf{p}} \to
\mathbf{p}+\tau_3 \mathbf{p}_s$ and $e\varphi \to \Phi$ in the
dynamical equation (\ref{eqWDF}), as well as in the definition of
the current in Eq.~(\ref{j1}). Note that the anisotropic term
$\mathbf{p \cdot v}_s$ arising from $\tilde{\mathbf{p}}$ in
Eq.~(\ref{eqWDF}) commutes with $\widehat{f}$ and thus drops out
from this equation.

While the physical sense of ${\bf p}_s$ is obvious, the
interpretation of the gauge invariant potential $\Phi$ is less
evident. Within the framework of the two-fluid model, it can be
interpreted as the difference $\Phi = \mu_s - \mu_n$ between the
electrochemical potentials of the condensate of Cooper
pairs, $\mu_s = \mu + (1/2)\partial\chi /\partial t$ and quasiparticles $%
\mu_n = \mu - e\varphi$; thus, a nonzero value of $\Phi$ means the
nonequilibrium of the electrons in the superconductor. In bulk
superconductors, $\Phi$ and ${\bf p}_s$ decay within their
corresponding lengths: London (skin) depth $\delta$ for
$\mathbf{p}_s$ and the electric field penetration depth
$\lambda_E$ for $\Phi$.

\section{Wigner distribution function for homogeneous superconducting systems%
}

In what follows, we focus on homogeneous superconducting systems
in pure limit, assuming the scattering rate is much smaller than
$\Delta$. To be more specific, we assume the magnitude of the
order parameter $\Delta$ and the gradient of its phase, $\nabla
\chi $, to be uniform inside the superconductor. Using an
appropriate gauge transformation, we include the
spatially varying part of the phase of $\Delta $ into the homogeneous $%
\mathbf{p}_{s}$. A ``residual'' spatially uniform phase is kept to
describe the dynamics of the phase of the order parameter. It can
be related to, e.g., possible (time-dependent) phase on either
sides of a Josephson junction. In this case, the equation for the
WDF takes the form,
\begin{equation}
\frac{\partial \widehat{f}}{\partial t}+i[\tilde{\xi}_{p}\tau _{3}-\widehat{%
\Delta },\widehat{f}] + \nu (\widehat{f}-\widehat{f}_0)=0.  \label{homeq}
\end{equation}
where $\tilde{\xi}_{p}=\xi _{p}+\Phi +mv_{s}^{2}/2$ and $\xi_p = {\ p}%
^2/2m-\mu$. The phenomenological collision term $\nu (\widehat{f}-\widehat{f}%
_0)$ qualitatively describes slow relaxation of the WDF to its equilibrium
value $\widehat{f}_0$ which is associated with the interaction Hamiltonian $%
H_1$. In the collisionless limit considered below, we will assume $\nu \to
+0 $, in order to provide correct analytical behavior of the WDF at $t \to
+\infty$.

Equation (\ref{homeq}) has several important properties which can be derived
from the equations for the matrix elements of $\widehat{f}$,
\begin{align}
i{\frac{\partial f_{11}}{\partial t}} &=-i{\frac{\partial f_{22}}{\partial t}%
}=-(\Delta f_{21}-\Delta ^{\ast }f_{12}),  \label{eqf11} \\
i{\frac{\partial f_{12}}{\partial t}} &=2\tilde{\xi}_{p}f_{12}+\Delta
(f_{11}-f_{22}),  \label{eqf12} \\
-i{\frac{\partial f_{21}}{\partial t}} &=2\tilde{\xi}_{p}f_{21}+\Delta
^{\ast }(f_{11}-f_{22}).  \label{eqf21}
\end{align}
First, we note that only the difference $f_{11}-f_{22}$ of the
diagonal components of the matrix $\widehat f$, enters the
equations for the off-diagonal components $f_{12}$ and $f_{21}$.
Furthermore, from Eq.~(\ref{eqf11}),
one finds that the sum of the diagonal components $%
f_{11}+f_{22}= \mbox{const}$. This allows us to present the function $%
\widehat{f}$ in the following form:
\begin{equation}
\widehat f = {\frac{1 }{2}} \left[ \widehat{1} (1 - \mathcal{F}_-) -
\widehat{\tilde f} \mathcal{F}_+ \right], \qquad \widehat{\tilde f}= \left(
\begin{array}{cc}
-g & f \\
f^* & g
\end{array}
\right),  \label{ftld}
\end{equation}
where $f$ and $g$ are isotropic functions, and the
time-in\-de\-pen\-dent quantities $\mathcal{F}_\pm$ have the
meaning of quasiparticle distribution functions which are
conserved during the stage of the collisionless evolution.
Assuming the system to be initially in equilibrium and comparing
Eq.~(\ref{ftld}) with the equilibrium form of the WDF, which can
be directly obtained from the definition in Eq.~(\ref{WDF})
\begin{equation}
\widehat f_0 = {\frac{1 }{2}} \left\{\widehat{1} \left( 1 - \mathcal{F}_-
\right) - {\frac{1 }{\tilde \epsilon_p}} \left( \tilde \xi_p \widehat \tau_3
- \widehat \Delta \right) \mathcal{F}_+ \right\},  \label{eqcond}
\end{equation}
we find the distribution functions
\begin{equation}
\mathcal{F}_\pm = {\frac{1 }{2}} \left[ \tanh \frac{\tilde \epsilon_p +
\mathbf{p \cdot v}_s(0) }{2T } \pm \tanh \frac{\tilde \epsilon_p - \mathbf{p \cdot v}%
_s(0) }{2T } \right], \quad \tilde \epsilon_p = \sqrt{\tilde \xi_p^2 +
|\Delta|^2},
\end{equation}
and the equilibrium values of the functions $f$ and $g$
\begin{equation}
f_0 = {\frac{\Delta }{\tilde \epsilon_p}}\ , \qquad g_0 = {\frac{\tilde
\xi_p }{\tilde \epsilon_p}}.  \label{fg0}
\end{equation}

In this representation, the dynamic equation (\ref{homeq}) for the
WDF reduces to the following system of scalar equations for the
functions $g$ and $f$:
\begin{align}
{\frac{\partial g }{\partial t}} &= i(\Delta^* f - \Delta f^*),  \notag \\
{\frac{\partial f }{\partial t}} &= 2i(\Delta g- \tilde \xi_p f ) ,
\label{homeq1}
\end{align}
which, together with Eq.~(\ref{fg0}), lead to the normalization condition
\begin{equation}
g^2+ff^*= 1.  \label{norm}
\end{equation}
The self-consistency equation has the form
\begin{equation}
\Delta(t) = \frac{\lambda}{2} \int {\frac{d\Omega_{p} }{4 \pi}}
\int_{-\omega_D}^{\omega_D} d\xi_{p} f(\xi_p,t)\mathcal{F}_{+},
\label{Delta2}
\end{equation}
where $\omega_D$ is the Debye frequency, $\lambda= N(0)V_0$ is the
dimensionless pairing constant, $N(0)$ is the electron density of states per
spin at the Fermi level, and $\Omega_p$ denotes angle variables associated
with the momentum vector. The charge and current densities are given by
\begin{align}
\rho(t) &= -e N(0) \int {\frac{d\Omega_p }{4 \pi}} \int d\xi_{p} \,
g(\xi_{p},t) \mathcal{F}_{+},  \label{rho2} \\
\mathbf{j}(t) &= en\mathbf{v}_s(t) -e N(0) \int {\frac{d\Omega_p }{4 \pi}} \,%
\mathbf{p}\int d\xi_{p}\, \mathcal{F}_-,  \label{j2}
\end{align}
where $n$ is the net electron density. Equation (\ref{j2}) shows that the
electric current is governed directly by the superfluid velocity and has
nothing to do with the evolution of the WDF,
\begin{equation}  \label{j3}
{\mathbf{j}}(t) = en{\mathbf{v}}_s(t) + e(n_{s} -
n){}\mathbf{v}_{s}(0) = {\mathbf{j}}(0) +
en[{\mathbf{v}}_s(t)-{\mathbf{v}}_s(0)],
\end{equation}
where $n_s$ is the condensate density calculated for the initial
superfluid velocity ${\mathbf{v}}_{s}(0)$. This property reflects
the specifics of the collisionless regime, in which the normal
component of the current flow is not affected by scattering, and
therefore the velocities of both the superfluid and normal
components of the electron fluid undergo equal changes
${\mathbf{v}}_s(t)-{\mathbf{v}}_s(0)$: ${\mathbf{v}}_s(0) \to
{\mathbf{v}}_s(t)$, ${\mathbf{v}}_n(0)=0 \to
{\mathbf{v}}_s(t)-{\mathbf{v}}_s(0)$. From this we conclude that
at nonzero temperature, when the density of the normal component,
$n_{n} \equiv n-n_{s}$, is nonzero, the current reverses its
direction with respect to ${\mathbf{v}}_s(t)$ if the latter
becomes smaller than ${\mathbf{v} }_{s}(0) n_{n}/n$.

\section{Collisionless evolution of the order parameter in superconductors.}

In the paper by Volkov and Kogan \cite{volkov}, the problem of
evolution of the order parameter $\Delta (t)$ at given initial
value of the WDF (and corresponding initial self-consistent value
of $\Delta =\Delta (0)$) was analyzed within a
linear approximation, assuming small deviations of $\Delta (t)$ and $\widehat{f}%
(\xi ,t)$ from their equilibrium values . It was shown that the
time variations of $\Delta $ have the form of harmonic
oscillations with the period of the order of $\Delta^{-1} $ and
the amplitude slowly decreasing as $t^{-1/2}$. At large $t\gg t_0
= \Delta^{-1}(0)$, the order parameter approaches a constant value
$\Delta _{\infty }\equiv \Delta (t\rightarrow \infty )$, which is
determined by the initial conditions and coincides neither with
$\Delta (0)$, nor with the equilibrium value $\Delta _{0}$.

In this paper, we address a more general nonlinear problem, with
arbitrary initial conditions, which may essentially differ from
the equilibrium state. In particular, this allows us to consider
formation of the superconducting state from the initial normal
state at low enough temperatures, or destruction of the initial
superconducting state at high temperatures. To this end, we apply
a numerical procedure, by making use of the so-called Riccati
parametrization of the functions $f$ and $g$. Due to the
normalization condition (\ref{norm}), these functions can be
expressed through a single function $a(\xi _{p},t)$,
\begin{equation}
g={\frac{1-aa^{\ast }}{1+aa^{\ast }}},\qquad f=\frac{2a}{1+aa^{\ast}},
\end{equation}
which satisfies a nonlinear Riccati-type equation,
\begin{equation}  \label{riccati}
{\frac{\partial a}{\partial t}}=i\left(-2\tilde{\xi}_{p}a-\Delta ^{\ast
}a^{2}+\Delta \right).
\end{equation}
In the stationary limit ($\Delta=\mathrm{const}$), the solution of Eq.~(\ref
{riccati}) is
\begin{equation}
a_{0}={\frac{\Delta }{\tilde{\xi}_{p}+\tilde{\epsilon}_{p}}}.
\end{equation}

\begin{figure}[h]
\includegraphics[width=9cm]{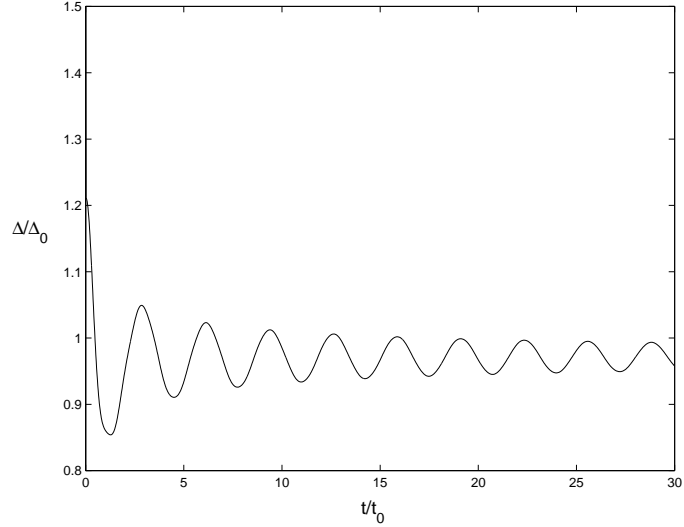}
\caption{ Collisionless time evolution of \ the order parameter
with initial value $\Delta (0)$ larger than the equilibrium value
$\Delta _{0}$ at T=0. In all figures the time is normalized on
$t_0=1/\Delta_0$.}
\end{figure}

In a general non-stationary case, one needs to integrate Eq.~(\ref{riccati}%
), together with the self-con\-sistency equation (\ref{Delta2}). Thus,
proceeding to the discrete time variable, $t =n \delta t$, $n=0,1,\ldots$,
one has to calculate the new value of $\Delta $ from Eq.~(\ref{Delta2})
after each time step $\delta t$, and then use it for the next step. For
sufficiently small $\delta t$, $\Delta$ can be approximately considered as
constant between $t$ and $t+\delta t$, which allows us to apply an
analytical solution of Eq.~(\ref{riccati}) within this time interval,
\begin{equation}
a(t+\delta t)=a(t)+{\frac{\Delta (t)-2\tilde{\xi}_{p}a(t)-\Delta ^{\ast
}(t)a^2(t)}{\Delta ^{\ast }(t)a(t)+\tilde{\xi}_{p}-i\tilde{\epsilon}_{p}\cot
(\tilde{\epsilon}_{p}\delta t)}},  \label{at}
\end{equation}
and thus to calculate $a(t+\delta t)$ explicitly. As the result, the
numerical procedure reduces to the numerical solution of the
self-con\-sistency equation at each step of calculations.

\begin{figure}[h]
\includegraphics[width=9cm]{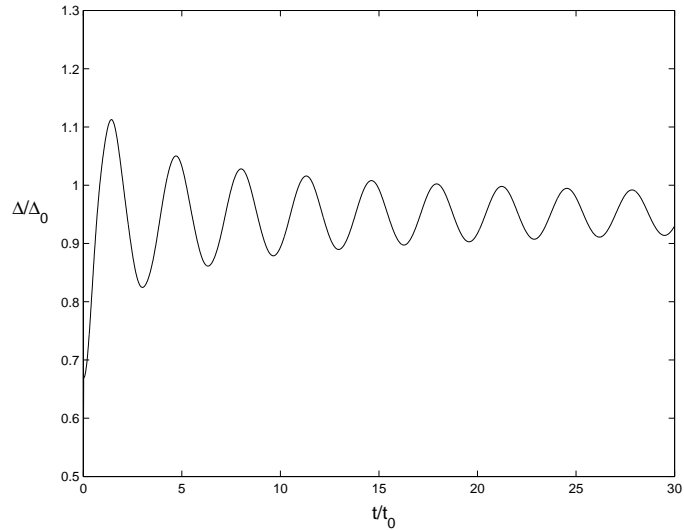}
\caption{Collisionless time evolution of \ the order parameter
with initial value $\Delta (0) < \Delta_0$ at $T=0$.}
\end{figure}

In our calculations, we use time steps $\delta t=0.02t_{0}$. After
each step, the values of the modulus and the phase of $\Delta (t)$
were recalculated by means of the self-consistency equation (31).
In Figs.~1 and 2, we present time variations of the order
parameter modulus with the initial values $\Delta (0)$ essentially
different from the equilibrium value $\Delta _{0}$ at $T=0$. It is
obvious that equal values of $\Delta (0)$ may be obtained for
different forms of the initial Wigner distribution function
$\widehat{f}(0)$. In our evaluation, we use the equilibrium form
of $\widehat{f}(0)$ given by Eq.~(\ref{eqcond}) at $T=0$, with a
formal parameter $\Delta _{\mathrm{in}}$, which, however, appears
to be slightly different from the initial self-consistent value
$\Delta (0)$. This difference weakly depends on the value of the
pairing constant $\lambda$, which we in the following put
$\lambda=0.5$. The initial value of $\Delta
_{\mathrm{in}}=1.5\Delta _{0}$ leads to $\Delta (0)\approx
1.3\Delta _{0}$ (Fig.~1), whereas $\Delta _{\mathrm{in}}=0.5\Delta
_{0}$ yields self-consistent $\Delta (0)\approx 0.67\Delta _{0}$
(Fig.~2).

Another type of perturbation in the system is the switching of the
$\lambda$ from one value to another. Or, more generally, the case
of time dependent BCS pairing. We have used the equations (35),
(36), (38) and (31) (with $\lambda=\lambda (t)$) to study this
problem numerically. In Fig.3 the collisionless evolution of the
order parameter under the changing of $\lambda$ in time is shown.

\begin{figure}[h]
\includegraphics[width=9cm]{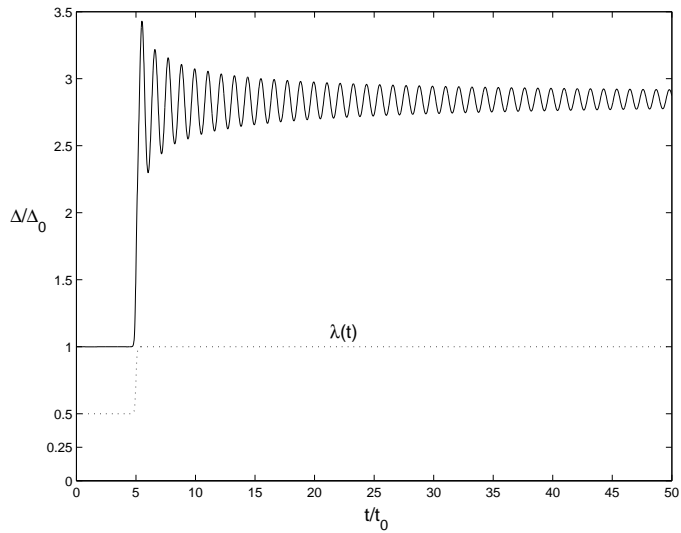}
\caption{Collisionless time evolution of the order parameter under
the changing of coupling constant $\lambda $\ from the value $0.5$
to the $1$.}
\end{figure}

It is interesting to note that the initial BCS form of the WDF
automatically leads to conservation of arbitrary initial value of
the order parameter phase $\chi$. Actually, such property is
associated with the definite symmetry of the initial WDF with
respect to $\xi_{p}$, which holds during the time evolution,
$f(\xi _{p},t)=f(-\xi _{p},t)$, $g(\xi _{p},t)=-g(-\xi _{p},t)$,
and manifests equality of the populations of the electron- and
hole-like excitations with equal energies $\tilde{\epsilon}_{p}$.
The introduction of an imbalance between the electron and hole
branches of the excitation spectrum (i.e., violation of the
above-mentioned symmetry) produces an excess charge in the
quasiparticle subsystem which, due to electroneutrality of the
metal, should be compensated by the opposite charge of the
superfluid condensate. This means the appearance of the difference
$\delta \mu $ between the electrochemical potentials $\mu _{n}$
and $\mu _{s}$ of excitations and the condensate, respectively,
which produces time variations of the order parameter phase
according to the relationship $d\chi /dt=2\delta \mu $. For a
given constant $\delta \mu $, we find continuous variation of the
phase with a constant rate.

The processes of formation and destruction of the superconducting
state can be also analyzed within the nonlinear collisionless
approach. By starting evaluations from a very small value of
$\Delta _{in}$ ($\sim 10^{-3}\Delta _{0}$) in Eq.\ (\ref{fg0}) at
$T=0$, which approximately represents initial normal state (see
Fig.~4, we observe a rapid increase in $\Delta (t)$ at the time
$t\sim t_{0}$ up to $\Delta \sim \Delta _{0}$, followed by an
oscillatory approach to a stable superconducting state. We note
that the asymptotic value $\Delta _{\infty }$ appears to be
noticeably lower than $\Delta _{0}$, which means that the real
equilibrium value of $\Delta$ at the superconducting transition is
formed via the relaxation processes.

\begin{figure}[h]
\includegraphics[width=9cm]{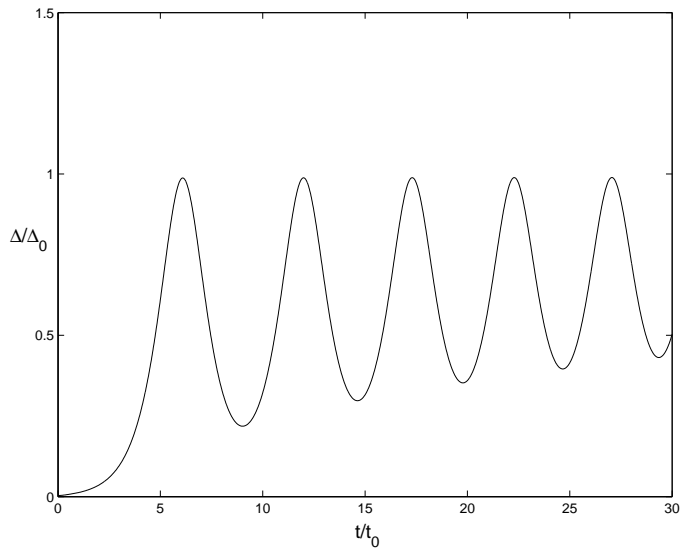}
\caption{Instability of the equilibrium normal state at $T=0$. We
start from $\Delta (0)=0.001\Delta _{0}$.}
\end{figure}

Strictly speaking, at any temperature, including the region
$T<T_c$, the self-consistency equation (\ref{Delta}) always has a
trivial solution $\Delta=0$, which corresponds to the normal
state. However, at $T<T_c$, the normal state is associated with a
maximum of the free energy, and therefore Fig.~4 actually
illustrates the thermodynamic instability of the normal state with
respect to an infinitesimal $\Delta$, which develops through the
quantum kinetic process described by Eqs.~(\ref{homeq1}) and
(\ref{Delta}). It is interesting to note that, despite the strong
nonlinearity of the process, the oscillations of $\Delta(t)$ have
almost purely harmonic shape.

The instability of the superconducting state at the temperature
$T>T_{c}$ is illustrated by Fig.~5, which was obtained by starting
evaluations from the initial superconducting state in
Eq.~(\ref{fg0}) at high enough temperature $T=2.5\Delta _{0}$. The
order parameter decreases approximately exponentially with the
characteristic decay time $0.42t_{0}$ without any oscillations. At
the final stage of the evolution, the order parameter enters the
fluctuation regime which is out of the framework of our
self-consistent approach.

\begin{figure}[h]
\includegraphics[width=9cm]{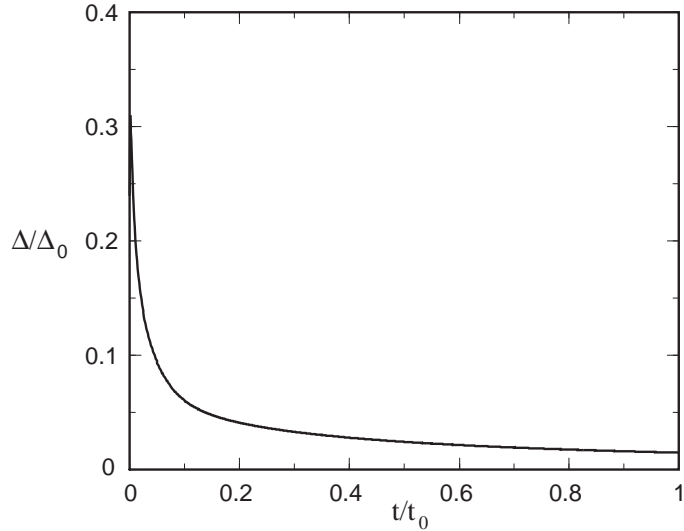}
\caption{Instability of the superconducting state at $T=2.5\Delta
_{0}>T_{c}$. Initial condition $\Delta (0)=0.31\Delta _{0}$.}
\end{figure}

In conclusion, the authors are grateful to B.Z. Spivak and A.M.
Zagoskin for discussing problems encountered in this work.

\bigskip

\bigskip


\bigskip


\end{document}